# Long lived surface plasmons on the interface of a metal and a photonic time-crystal

Lior Bar-Hillel[1+], Yonatan Plotnik[2], Ohad Segal[1] and Mordechai Segev[1,2]

1. *Department of Electrical and Computer Engineering, Technion, Haifa 32000, Israel*
2. *Physics Department, Technion, Haifa 32000, Israel*

**Abstract**

We predict the existence of surface plasmons polaritons at the interface between a metal and a periodically modulated dielectric medium, and find an unusual multi-branched dispersion curve of surface and bulk modes. The branches are separated by momentum gaps indicating intense amplification of modes, and display high and low group velocity ranging from zero to infinity at short wavelengths. We simulate how these SPP modes are formed by launching a properly engineered laser beam onto the metallic interface and examine their space-time evolution. The amplification of the surface plasmons at the interface with a photonic time-crystal offers a path to overcome plasmonic losses, which have been a major challenge in plasmonics.

Recent years have witnessed a surge of research interest of light in time-varying media and specifically in Photonic Time-Crystals (PTCs) [1–6]. This emerging field offers intriguing insights into the fundamental nature of time in wave systems, especially in the context of light-matter interactions. PTCs are materials that exhibit a large and periodic change in their electromagnetic properties, occurring within just a few cycles of the wave propagating within them (Fig 1(a)). Each abrupt change in the properties of the medium gives rise to time-refracted and time-reflected waves, which interfere and result in a distinctive band structure in the dispersion relation. The dispersion relation of a PTC displays momentum bands with real frequencies, separated by momentum gaps where the frequency is complex, Fig 1(b). While PTCs bear some resemblance to one-dimensional spatial photonic crystals in terms of their dispersion relations, they also display profound fundamental differences. In PTCs, breaking the time-translation symmetry (due to the temporal modulation) results in the absence of energy (frequency) conservation, unlike spatial photonic crystals that conserve energy but not momentum. Thus, the amplitudes of the modes within the momentum gaps of PTCs can also undergo exponential growth in time, extracting energy from the modulation. It is this momentum gap which makes PTCs so special, especially for light-matter interactions, on both the classical and the quantum levels. For example, a subwavelength dipole radiating at an arbitrary frequency always has some wave-momentum components associated with the gap, which are amplified by the modulation and become more and more coherent as time lapses [7]. Likewise, an electron moving in the vicinity of a PTC emits radiation even below the Cherenkov threshold, and displays quantum interference effects [8]. However, Observing the distinctive features of PTCs, such as momentum gaps, requires a significant modulation amplitude at a period comparable to a single cycle of the EM wave propagating within the medium. These requirements are extremely hard to meet, which is why thus far PTCs have been demonstrated only at radio frequencies [9], and time-reflection was observed only recently and in microwaves [10,11] but never at optical frequencies. Nevertheless, recent advances with transparent conductive oxides, especially the recent observation of single-cycle time-refraction and extremely fast relaxation [12,13], hold the promise of realizing PTCs at optical frequencies in the near future. These advancements pave the way for exciting experiments and further exploration of PTCs in the optical regime.

An electromagnetic (EM) phenomenon that can highly benefit from the amplification in PTCs is surface plasmon polaritons (SPPs). Surface plasmons polaritons arise from strong coupling between EM waves and collective oscillations of free electrons at metal-dielectric interfaces [14,15]. Plasmons enable the confinement of EM energy to subwavelength dimensions, surpassing the diffraction limit of conventional optics. Through their remarkable ability to confine and manipulate light on nanoscales, SPPs have paved the way for advances in areas such as sensing, spectroscopy, imaging, and data communication. However, despite their promising potential, SPPs encounter two fundamental limitations rooted in metal losses, which restrict their practical applications. The first limitation concerns the minimum achievable wavelength, while the second is a limited propagation distance before their energy dissipates [16,17]. In this work, we leverage the unique band structure of a PTC to overcome those inherent limitations.

Here, we investigate surface plasmon polaritons (SPPs) on the interface between a metal and a time-varying medium, focusing on a metal-PTC interface. We find their dispersion curve and their wavefunctions, identify the regimes of amplification associated with the momentum gaps where the SPPs draw energy from the modulation, and pinpoint regions of slow and fast light at short wavelengths. Importantly, the amplification of the SPPs at the metal-PTC interdace can overcome plasmonic losses, offering an avenue for using longer range plasmonic waveguides and related devices.

Consider an interface between a PTC and a metal at the plane $z = 0$, where the PTC occupies the half space $z > 0$ and the metal fills the rest, as illustrated in Fig. 1(c). To characterize the permittivity of the metal, which we denote by $\varepsilon_m(\omega)$, either the Drude model or the Lorentz oscillator model can be employed. The EM modes within the metal are described by the dispersion relation $k_m(\omega) = \omega\sqrt{\mu_0 \varepsilon_m(\omega)}$. The PTC is characterized by a periodic time-dependent permittivity $\varepsilon(t) = \varepsilon(t + T)$ with a modulation frequency $\Omega = 2\pi/T$. Thus, The EM fields propagating within the PTC are solutions of the wave equation $(\mu_0 \partial_t(\varepsilon(t)\partial_t) - \vec{\nabla})\vec{H}(\vec{r}, t) = 0$. Applying the Bloch-Floquet theorem, we find a TM solution of the form $\vec{H}(\vec{r}, t) = \left(\sum_{n=-\infty}^{\infty} h_n(\omega) e^{-in\Omega t}\right) e^{i(\vec{k}\cdot\vec{r} - \omega t)} \hat{y}$. Substituting this solution into the wave equation leads to eigenmodes with coefficients $h_n(\omega)$, exhibiting the band structure of the PTC given by the eigenvalues $k_{PTC}(\omega)$ (Fig. 1(b)). We look for SPP modes on the interface between the metal and the PTC, with frequency $\omega$ and propagation constant (momentum) $k_{spp}$ (Fig. 1(c)). Since the PTC is formed by a periodic modulation at frequency $\Omega$, an excitation at frequency $\omega$ is coupled by the PTC to modes with frequencies $\omega \pm n\Omega$, for any integer $n$, which in turn excites plasmons with those frequencies in the metal. Thus, such SPP mode must contain an infinite discrete set of wavevectors and frequencies. The most general form of the magnetic field is obtained by sum over the different bands of the PTC

$$H_y(\vec{r}, t) = \sum_{b=1}^{\infty} A_b(\omega) \sum_{n=-\infty}^{\infty} h_{bn}(\omega) e^{i(k_{spp}x - (\omega + n\Omega)t)} \begin{cases} e^{-\kappa_{b,PTC}z}, & z \geq 0 \\ e^{\kappa_{n,m}z}, & z < 0 \end{cases} \quad (1)$$

Where $A_b$ is the amplitude of the mode at band $b$. Imposing the boundary conditions on the transverse electric and magnetic fields (i.e. their continuity), we find a set of coupled equations for the amplitudes $A_b$ which depend on the integer $n$

$$\sum_{b=1}^{\infty}\left(\frac{\sqrt{k_{spp}^2 - k_{PTC,b}^2(\omega)}}{k_{PTC,b}^2(\omega)} + \frac{\sqrt{k_{spp}^2 - k_m^2(\omega + n\Omega)}}{k_m^2(\omega + n\Omega)}\right) h_{bn}(\omega) A_b(\omega) = 0. \tag{2}$$

This relation holds for any integer $n$. Derivation of Eq. 2 is presented in section B of [18]. This set of equations can be arranged in a matrix form $M(\omega, k_{spp})\vec{A}_b(\omega) = 0$. Thus, by evaluating the determinant of the matrix $M$, we find all the values of $\omega, k_{spp}$ for which nontrivial solutions exist, and their wave functions. Figure 2 shows $k_{spp}(\omega)$ for the case of a lossless metal, which we model by free electron gas such that $k_m(\omega) = \frac{\omega}{c}\sqrt{1 - \left(\frac{\omega_p}{\omega}\right)^2}$. To address real metals with energy dissipation, we employ the Drude model as explained in the conclusions section and presented in section C of [18].

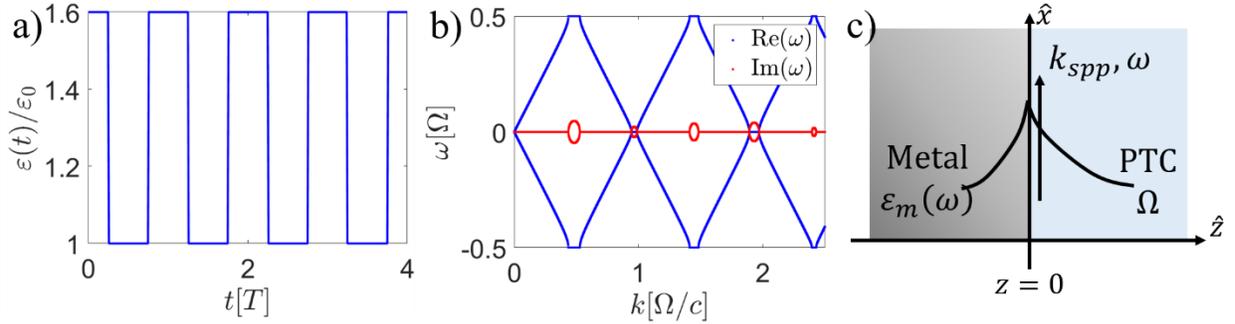

Fig. 1: **Metal-PTC interface.** (a) The relative permittivity giving rise to the PTC as a function of time. (b) Band structure of the PTC with the permittivity of (a). (c) An interface between a metal and a PTC

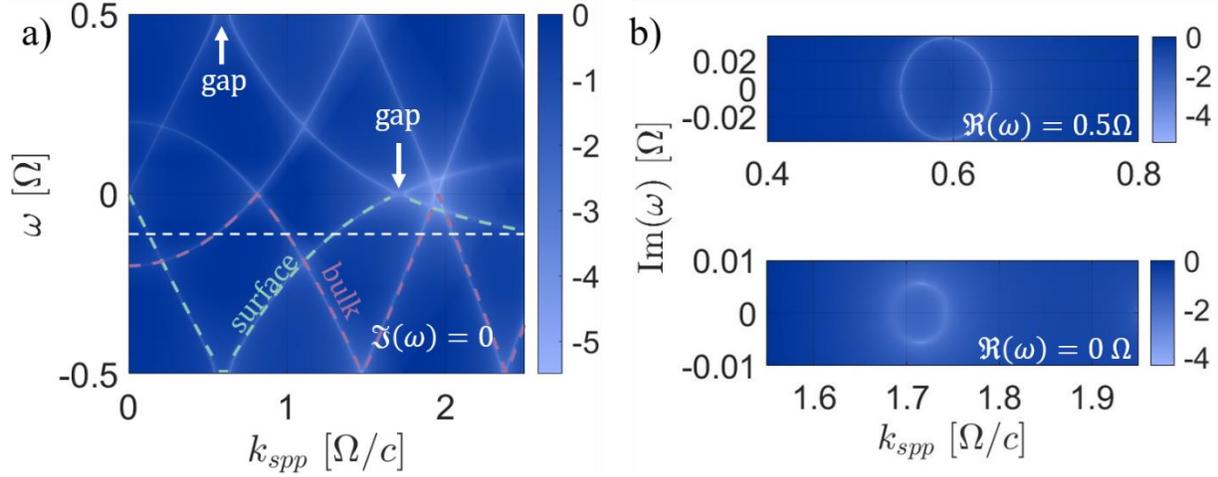

Fig. 2: **Band structure of SPPs at a metal-PTC interface.** (a) Band structure of SPPs for $\omega_p = 1.8\Omega$. The color denotes the value $\det M(\omega, k)$, which is dimensionless, in logarithmic scale. A solution to Eq. (2) exists along the white curves which form the dispersion relation. (b) Momentum bandgaps of the band structure presented in (a).

Examining Fig. 2, we see that the dispersion curve of the SPPs displays several bulk modes bands and a several surface modes bands. Increasing the ratio $\omega_p/\Omega$ will increase the number of surface mode bands. The last band of the surface modes have a maximum frequency, marked by a white line in Fig. 2(b). The surface bands are separated by momentum gaps which host EM modes with complex Floquet frequencies, giving rise to exponential amplification (or decay) with time. The real value of the frequency within the gap is constant, resulting in a group velocity of zero, so pulses within the gaps are stationary, i.e., they do not propagate. These SPPs exhibit additional unique features. For example, we identify points on the dispersion curve of surface modes which display extremely high group velocities close to bandgap. Moreover, the SPP mode expressed by Eq. (1) contains an infinite, unbounded, discrete set of frequencies $\omega \pm n\Omega$, but it is associated with a single finite propagation constant, $k_{spp}(\omega)$ along the interface. At sufficiently high index $n$ where $k_{spp}(\omega) < k_m(\omega + n\Omega)$ the decay constant in the metal $\kappa_{n,m}$ becomes imaginary. Consequently, the corresponding Floquet mode, radiates in the -z direction, away from the interface, into the metal. The same mechanism leads to radiation in the z direction into the PTC away from the interface which comes from the higher bands. Thus, even though these modes are stationary – their shape and mean energy do not change with propagation – they radiate energy into the metal and into the PTC. This feature offers the ability to excite such SPP modes with a laser pulse, unlike ordinary SPPs at a metal-dielectric interface which have to be excited through a coupler [19,20]. These features enable new techniques for exciting SPPs and controlling their propagation. Notably, near the maximum frequency of the surface modes there are "slow light" modes which are a known feature of SPPs.

We procced with simulating the propagation of such SPPs, Fig. 3. In each simulation we create a simulation area of $3\mu m \times 20\mu m$, where the metal is in the region $0 \leq z \leq 0.5\mu m$, and a dielectric medium fills the rest. The plasma frequency of the metal is $\omega_p = 14 \cdot 10^{15} \frac{rad}{s}$. We excite a plasmonic pulse that propagates along the interface for a distance of $z = 0.5\mu m$. At time $t = 50 fs$, the PTC is turned on at the dielectric side, for the duration of 15 fs. The permittivity of the PTC is a square wave (as in Fig. 1) modulated between $\varepsilon_1 = 1$ and $\varepsilon_2 = 1.6$. The results are presented in Fig. 3 (a)-(f).

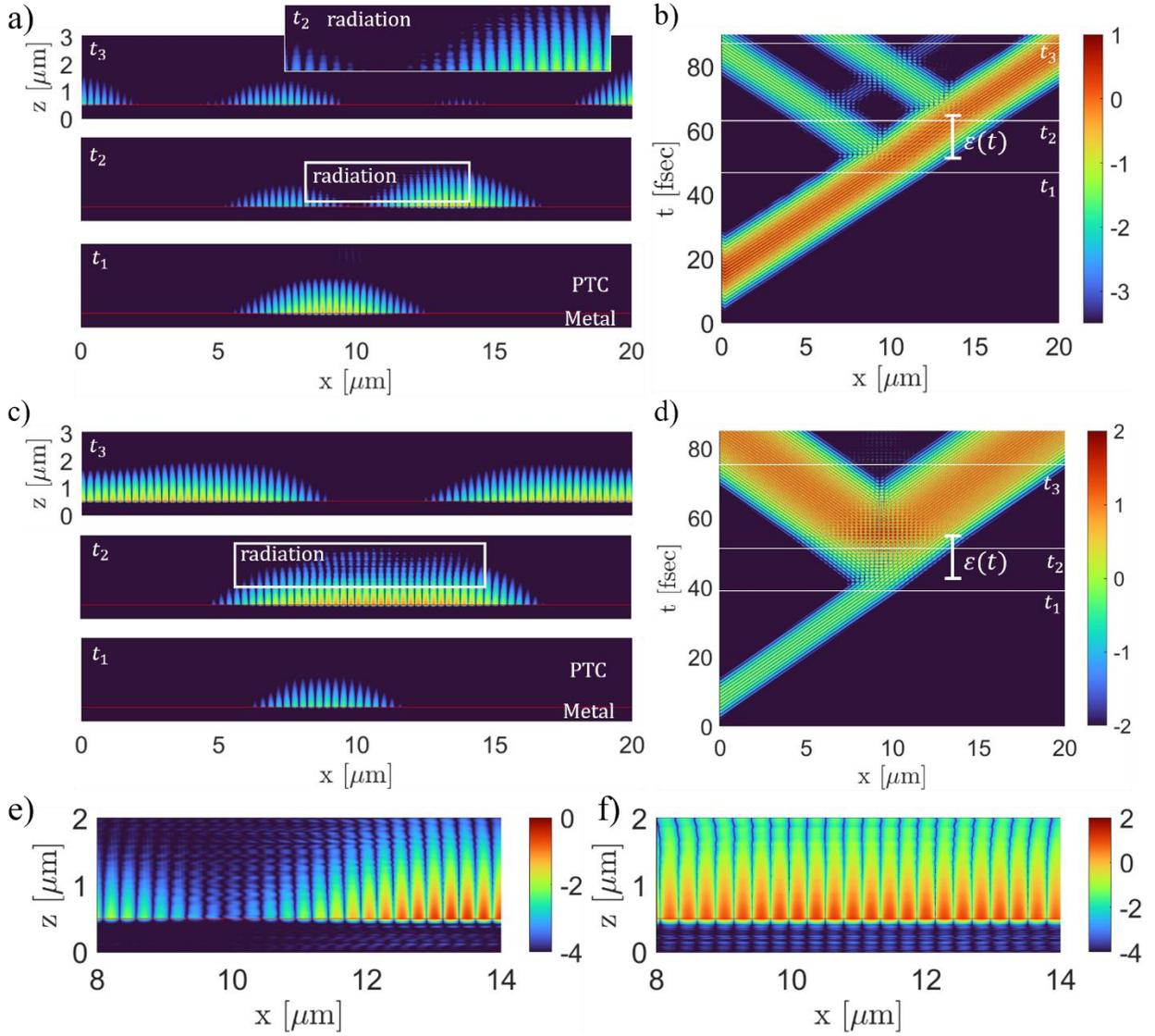

Fig. 3: **FDTD simulations of SPPs at a metal-PTC interface.** (a), (b) FDTD simulation for a PTC with a period $T = 1.5\ fs$. The magnitude of the magnetic field is presented in log scale (arbitrary units) with a common color bar. In this case, the mode is within a momentum band, and consequently there is no exponential amplification of the energy carried by the SPP. (c), (d) FDTD simulation for a PTC with a period $T = 1fs$. The magnitude of the magnetic field is presented in log scale (arbitrary units) with a common color bar. The plasmonic mode is within a gap, hence the energy is carried is exponentially amplified. The amplitude at the end of the PTC is 100 times larger than when the PTC has started. (a), (c) Show snapshots at 4 different times $t_1$-$t_3$. (b), (d) The amplitude of the magnetic field along the interface as a function of time ($H_z(z = 0.5\mu m, x, t)$). (e), (f) zoom in on the radiation of the SPP pulses into the metal and into to PTC at time. (e), (f) Zoom-in views of the SPP pulses at time $t_2$, corresponding to panels (a) and (c), respectively.

Figure 3(a,b) shows the evolution of an SPP for the case where the main wavenumber of the SPP pulse is within a momentum band. We observe that the turn-on of the PTC (at t=50fsec) splits the pulse into a forward propagating (time-refracted) SPP pulse and a backward propagating (time-

reflected) SPP pulse ($t_2$ panel in (a) and the corresponding first split in (b)). Likewise, when the PTC ends (at t=70fsec), each of the forward and backward propagating pulses splits again. At the end of the PTC, there are four SPP pulses propagating along the interface, as shown in Fig. 3(b). While the PTC is on, the SPP radiates into the dielectric medium and the metal, as shown by the interference fringes at time $t_2$ in Fig. 3(a) and 3(e). This radiation comes from the higher frequency components, generated by the modulation defining the PTC, as predicted according to Eq. (1) and (2). At the end of the PTC, the pulses are no longer radiating in the z direction, and indeed the interference fringes are absent at time $t_3$.

Figure 3(c,d) shows the evolution of an SPP for the case where the main frequency of the SPP pulse is within a momentum gap. In this case, at the moment the PTC turns on, the pulse stops and its amplitude starts to grow, but there is no splitting to forward and backward propagation for as long as the PTC is on – because the pulse is stationary (its group velocity is zero). When the PTC is turned off, the pulse splits into two SPP pulses, propagating in opposite directions along the interface. While the PTC is on, we see intense amplification, such that at the end of the PTC, the maximum intensity of the pulse is 100 times larger than the original intensity. Once again, while the PTC is on, the SPP radiates in the z direction into both the metal and the PTC, which can be noticed as the interference fringes at time $t_2$ in Fig. 3(c) and 3(f).

Next, we simulate the excitation of plasmonic modes using a laser pulse, Fig. 4. The pulse is propagating in a dielectric medium (from the bottom up), and is incident upon a metal strip of 100nm thickness, with a PTC at the other side. The PTC is the same as for Fig. 3, with a period of 1.5 fsec. As an example, we a excite a specific mode with negative group velocity, associated with the second branch. We choose the angle of incidence to be $\theta_i = 35^o$ (with the normal to the interface) and center frequency $f = 5.5 \cdot 10^{14} Hz$, for which the wavenumber in the z direction matches to a plasmon mode with the same frequency in the second branch. Due to the negative group velocity at this point in the dispersion curve, the plasmonic pulse propagates in a direction opposite to the excitation direction, as shown by the arrow in Fig. 4(c). The group velocity of the SPP and its direction (forward or backward) can be tuned by the modulation defining the PTC. As exemplified by this simulation, the PTC allows for new ways to control the propagation.

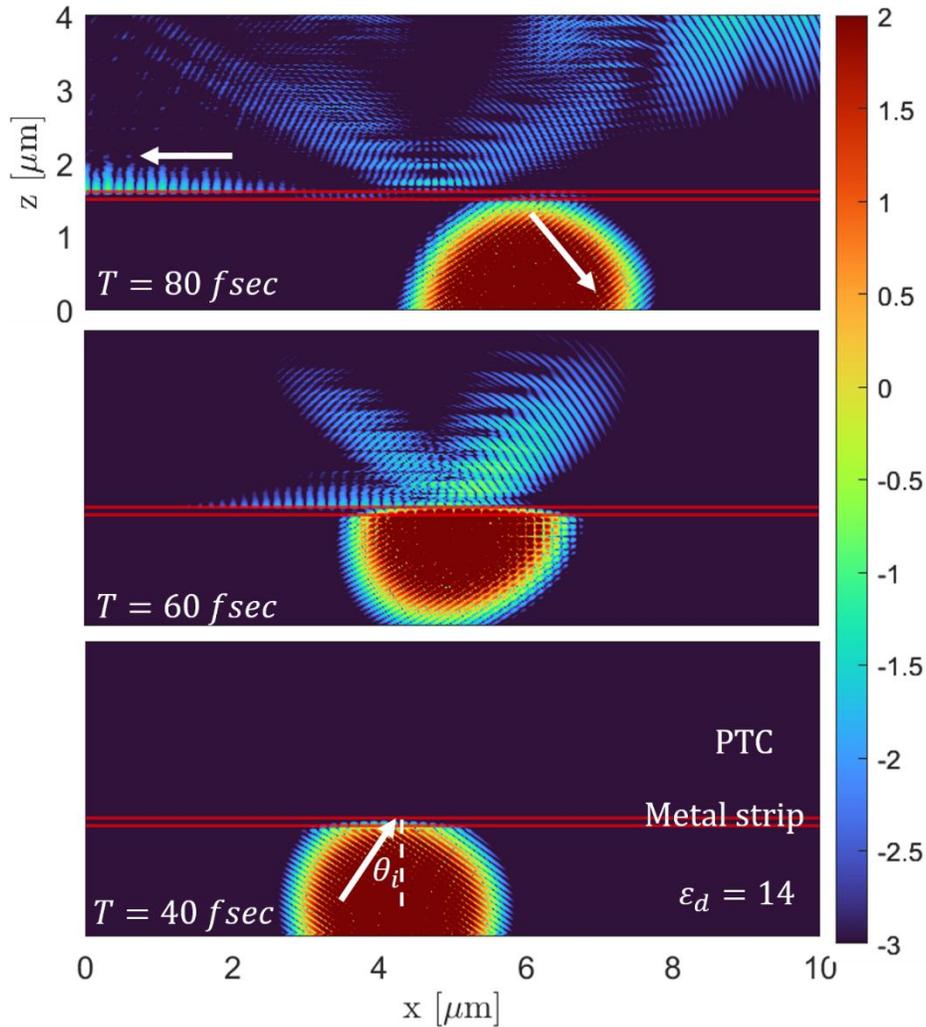

Fig. 4: **FDTD simulation of laser pulse excite SPPs at a metal-PTC interface.** A Gaussian pulse is incident from air onto a metal strip (marked between the two horizontal red lines) followed by a PTC (above the upper red line). The pulse excites an SPP mode with negative group velocity, and propagates along the metal-PTC interface in the backward direction, as indicated by the white arrow. The panels are ordered bottom to top and show consecutive snapshots in time of the SPP excitation and propagation.

In [18] we simulate a case for a non-ideal metal with energy dissipation and find that for a strong enough variation of the permittivity the amplification of the gap modes can overcome the intrinsic plasmonic losses. To highlight the feasibility of plasmonic time-crystals and their potential impact, we perform a simple estimation of the modulation strength and frequency required to overcome plasmonic losses. Assuming a modulation of the form $\varepsilon(t) = \varepsilon_1 + \varepsilon_2 \sin(\Omega t)$, the maximum value of the imaginary frequency is approximately $\omega_i \approx \frac{\Omega}{10}\frac{\varepsilon_2}{\varepsilon_1}$, based on empirical estimation from numerical calculations. The rate of energy dissipation in the metal is described by the parameter $\tau$ in the Drude model. To overcome the dissipation, $\tau < \omega_i^{-1}$ is required, which gives the inequality $\Omega\frac{\varepsilon_2}{\varepsilon_1} > \frac{10}{\tau}$. To that end, we note a recent work presenting refractive modulation of $\Delta n = 0.3$ in transparent conductive oxides, within a single optical cycle [12]. For a metal similar to silver, $\omega_p \approx 14 \cdot 10^{15} \frac{rad}{s}$ and $\tau \approx 15\ fsec$, we find that the modulation period should be $T \approx 7 fsec$ or shorter, to be able to overcome the metallic losses of SPPs. Thus, it is possible to use the amplification of SPPs to overcome plasmonic losses through the index modulation already with the current technology.

To conclude, we predicted the existence of surface plasmons polaritons on the interface between a metal and a photonic time-crystal, found their unique dispersion curve which consists of branches separated by gaps, and studies their propagation. We found that the modes residing within the momentum gaps grow exponentially by drawing energy form the modulation, and showed amplification which can overcome plasmonic losses, which are currently the main limitation of plasmonics. Finally, we find that the unique dispersion relation of these SPPs offers new ways to excite and control SPPs by shaping the modulation of the refractive index. These surface plasmons can offer new intriguing possibilities for interfaces between graphene (which supports plasmons in a wide range of frequencies [21–24]) and Moire superlattices [25] made from graphene (or other 2D materials) and a periodically modulated dielectric medium, especially given that the conduction properties in graphene can be tuned by electric fields. This actually suggests that plasmonic time-crystals can be tuned in real time by tuning the conduction properties – a fascinating vision in its own right.

**Note added during submission**: while submitting the article, related work has been brought to our attention [26]. Despite the related titles, that work has to do with bulk plasmons, not with surface plasmon polaritons.

# References

[1]     F. Biancalana, A. Amann, A. V. Uskov, and E. P. O'Reilly, Dynamics of light propagation in spatiotemporal dielectric structures, Phys. Rev. E **75**, 046607 (2007).
[2]     J. R. Zurita-Sánchez, P. Halevi, and J. C. Cervantes-González, Reflection and transmission of a wave incident on a slab with a time-periodic dielectric function ϵ(t), Phys. Rev. A **79**, 053821 (2009).
[3]     E. Lustig, Y. Sharabi, and M. Segev, Topological aspects of photonic time crystals, Optica, OPTICA **5**, 1390 (2018).
[4]     Y. Sharabi, E. Lustig, and M. Segev, Disordered Photonic Time Crystals, Phys. Rev. Lett. **126**, 163902 (2021).
[5]     E. Lustig, O. Segal, S. Saha, C. Fruhling, V. M. Shalaev, A. Boltasseva, and M. Segev, Photonic time-crystals - fundamental concepts [Invited], Opt. Express, OE **31**, 9165 (2023).
[6]     X. Wang, P. Garg, M. S. Mirmoosa, A. G. Lamprianidis, C. Rockstuhl, and V. S. Asadchy, Expanding momentum bandgaps in photonic time crystals through resonances, Nat. Photon. 1 (2024).
[7]     M. Lyubarov, Y. Lumer, A. Dikopoltsev, E. Lustig, Y. Sharabi, and M. Segev, Amplified emission and lasing in photonic time crystals, Science **377**, 425 (2022).
[8]     A. Dikopoltsev, Y. Sharabi, M. Lyubarov, Y. Lumer, S. Tsesses, E. Lustig, I. Kaminer, and M. Segev, Light emission by free electrons in photonic time-crystals, Proceedings of the National Academy of Sciences **119**, e2119705119 (2022).
[9]     J. R. Reyes-Ayona and P. Halevi, Observation of genuine wave vector (k or β) gap in a dynamic transmission line and temporal photonic crystals, Applied Physics Letters **107**, 074101 (2015).
[10]    H. Moussa, G. Xu, S. Yin, E. Galiffi, Y. Ra'di, and A. Alù, Observation of temporal reflection and broadband frequency translation at photonic time interfaces, Nat. Phys. **19**, 6 (2023).
[11]    T. R. Jones, A. V. Kildishev, M. Segev, and D. Peroulis, Time-reflection of microwaves by a fast optically-controlled time-boundary, Nat Commun **15**, 6786 (2024).
[12]    E. Lustig et al., Time-refraction optics with single cycle modulation, Nanophotonics **12**, 2221 (2023).
[13]    R. Tirole, S. Vezzoli, E. Galiffi, I. Robertson, D. Maurice, B. Tilmann, S. A. Maier, J. B. Pendry, and R. Sapienza, Double-slit time diffraction at optical frequencies, Nat. Phys. **19**, 7 (2023).
[14]    R. W. Wood, XLII. *On a remarkable case of uneven distribution of light in a diffraction grating spectrum*, The London, Edinburgh, and Dublin Philosophical Magazine and Journal of Science **4**, 396 (1902).
[15]    R. H. Ritchie, Plasma Losses by Fast Electrons in Thin Films, Phys. Rev. **106**, 874 (1957).
[16]    J. B. Khurgin and A. Boltasseva, Reflecting upon the losses in plasmonics and metamaterials, MRS Bulletin **37**, 768 (2012).
[17]    J. B. Khurgin and G. Sun, Practicality of compensating the loss in the plasmonic waveguides using semiconductor gain medium, Applied Physics Letters **100**, 011105 (2012).
[18]    See Supplemental Material, n.d.
[19]    E. Kretschmann and H. Raether, Notizen: Radiative Decay of Non Radiative Surface Plasmons Excited by Light, Zeitschrift Für Naturforschung A **23**, 2135 (1968).

[20] A. Otto, Excitation of nonradiative surface plasma waves in silver by the method of frustrated total reflection, Z. Physik **216**, 398 (1968).
[21] M. Jablan, H. Buljan, and M. Soljačić, Plasmonics in graphene at infrared frequencies, Phys. Rev. B **80**, 245435 (2009).
[22] M. Jablan, M. Soljačić, and H. Buljan, Plasmons in Graphene: Fundamental Properties and Potential Applications, Proceedings of the IEEE **101**, 1689 (2013).
[23] A. N. Grigorenko, M. Polini, and K. S. Novoselov, Graphene plasmonics, Nature Photon **6**, 749 (2012).
[24] J. Chen et al., Optical nano-imaging of gate-tunable graphene plasmons, Nature **487**, 77 (2012).
[25] G. X. Ni et al., Plasmons in graphene moiré superlattices, Nature Mater **14**, 1217 (2015).
[26] J. Feinberg, D. E. Fernandes, B. Shapiro, and M. G. Silveirinha, *Plasmonic Time Crystals*, arXiv:2407.19958.